\documentclass[sn-mathphys-num]{sn-jnl}

\usepackage{graphicx}%
\usepackage{multirow}%
\usepackage{amsmath,amssymb,amsfonts}%
\usepackage{amsthm}%
\usepackage{mathrsfs}%
\usepackage[title]{appendix}%
\usepackage{xcolor}%
\usepackage{textcomp}%
\usepackage{manyfoot}%
\usepackage{booktabs}%
\usepackage{algorithm}%
\usepackage{algorithmicx}%
\usepackage{algpseudocode}%
\usepackage{listings}%
\usepackage{cancel}
\usepackage{ulem}

\theoremstyle{thmstyleone}%
\theoremstyle{thmstyletwo}%

\theoremstyle{thmstylethree}%

\newcommand{\nn}{\nonumber\\}
\newcommand{\bec}[1]{\mbox{\boldmath $#1$}}
\newcommand{\XiN}{N\Xi}
\newcommand{\XiNN}{NN\Xi}
\newcommand{\XiNNN}{NNN\Xi}
\newcommand{\XiNa}{\alpha N\Xi}
\newcommand{\XiNaa}{\alpha\alpha N\Xi}
\newcommand{\Xiaa}{\alpha\alpha \Xi}
\newcommand{\XiMaa}{\alpha\alpha \Xi^-}

\begin{document}

\title[Article Title]{Cluster phenomena using few-body and Lattice QCD theories}


\author*[1,2]{\fnm{Emiko} \sur{Hiyama}}\email{hiyama@riken.jp}

\author[3,2]{\fnm{Takumi} \sur{Doi}}\email{doi@ribf.riken.jp}
\equalcont{These authors contributed equally to this work.}

\affil*[1]{\orgdiv{Department of Physics}, \orgname{Tohoku University}, \orgaddress{\street{Aoba 6-3}, \city{Sendai}, \postcode{980-8578}, \state{Miyagi}, \country{Japan}}}

\affil[2]{\orgdiv{Nishina Center}, \orgname{RIKEN}, \orgaddress{\street{Hirosawa 2-1}, \city{Wako}, \postcode{351-0198}, \state{Saitama}, \country{Japan}}}

\affil[3]{\orgdiv{RIKEN Center for Interdisciplinary Theoretical and
Mathematical Sciences (iTHEMS)}, \orgname{RIKEN}, \orgaddress{\street{Hirosawa 2-1}, \city{Wako}, \postcode{351-0198}, \state{Saitama}, \country{Japan}}}


\abstract{With the advancement of first-principles calculations for baryon-baryon interactions, 
it becomes possible to obtain reliable hyperon-nucleon potentials 
by lattice QCD simulations with the HAL QCD method.
High-precision few-body methods, such as the Gaussian Expansion Method (GEM), are applicable to 
solve quantum few-body systems up to
four- and five-body systems. By combining the HAL QCD potentials with the GEM, one can predict the level structure of novel hypernuclei prior to experimental observation.
In this review, we utilize the lattice QCD $N\Xi$ potential
obtained by the HAL QCD method
 to investigate the few-body systems $NN\Xi$ and $NNN\Xi$. Our analysis indicates that the lightest bound $\Xi$ hypernucleus is the $NNN\Xi$ system. To extract detailed information on the isospin and spin components  of the $N\Xi$ interaction, we perform a four-body calculation for the $\XiNaa$ system with the total isospin $T = 0$ and $T = 1$. We demonstrate that the level structure of this system is sensitive to the isospin and spin dependencies of the $N\Xi$ interaction.
Furthermore, we propose experimental investigations to produce the $NNN\Xi$ and $\XiNaa$ systems via the $(K^-, K^+)$ and $(K^-, K^0)$ reactions on $^4$He and $^{10}$B targets, respectively.}




\maketitle

\section{Introduction}\label{sec1}

It is  a fundamental question to understand how the hierarchical structure of quantum systems evolves.
Addressing this issue requires interdisciplinary collaboration among the fields of hadron physics, nuclear physics, atomic physics, and molecular physics.
Recent advancements in theoretical frameworks, along with the development of high-performance supercomputers such as "K" and "Fugaku," have enabled significant progress in ab initio calculations for rigorously solving quantum many-body systems. These developments have provided new insights into baryon-baryon interactions from first principles.
As a result, it has become possible to theoretically predict the level structures of novel  hypernuclei prior to experimental measurements, marking a major milestone in the study of quantum systems.

\begin{figure*}[htb]
\begin{center}
\includegraphics[scale=0.45]{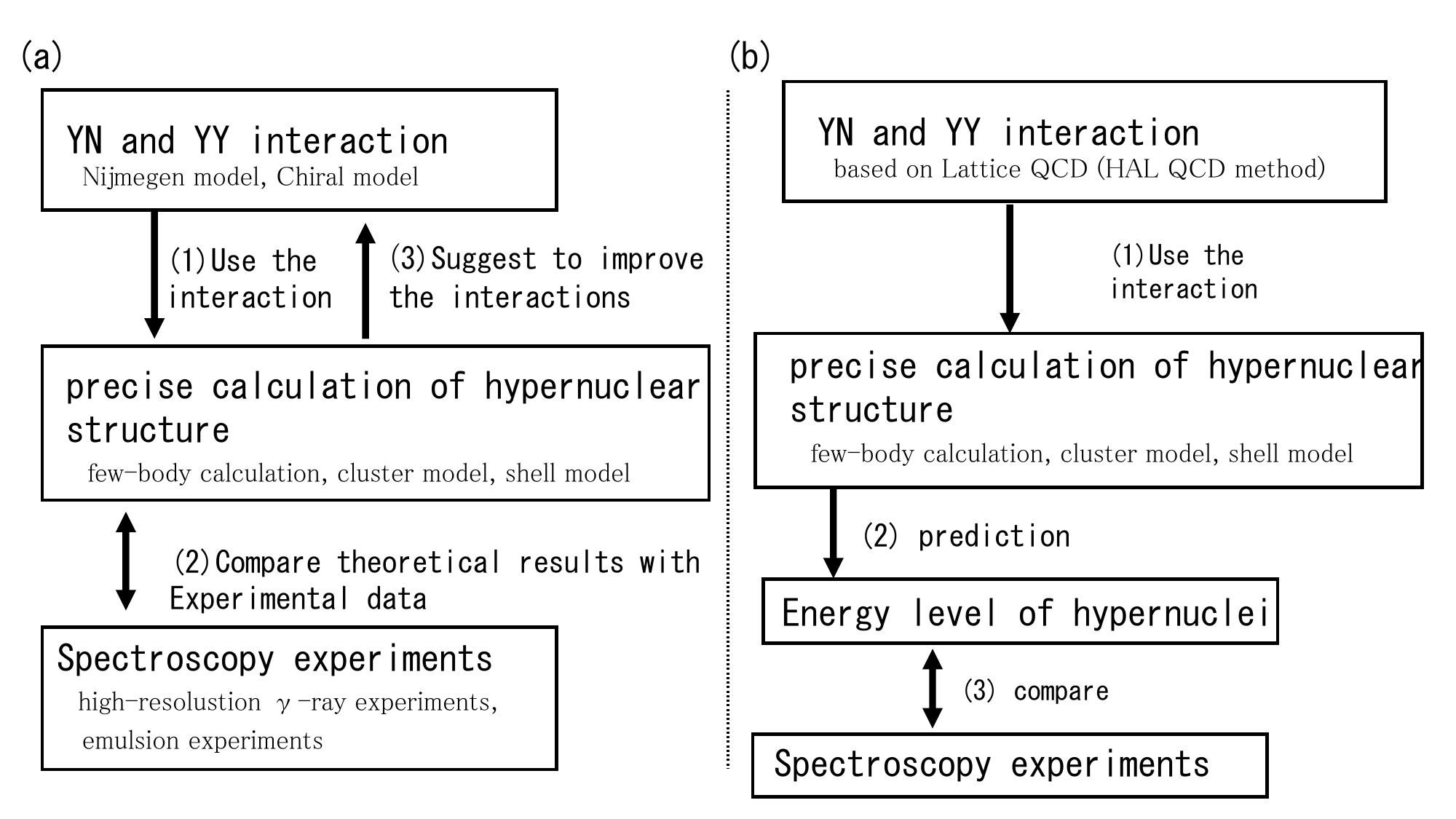}
\end{center}
\caption{
 The strategy for extracting information used so far is shown in (a),
    and the new strategy is shown in (b).
}
\label{fig:strategy}
\end{figure*}

Before the advancement of baryon-baryon interaction studies based on first-principles calculations, information had been obtained through the following strategy (Fig.~\ref{fig:strategy} (a)):
(i) Candidates for hyperon-nucleon ($YN$) and hyperon-hyperon ($YY$)  interactions were adopted, based on meson-exchange theory and/or chiral effective field theory.
(ii) Spectroscopy experiments of hypernuclei were utilized. In general, such experiments did not directly provide information on the $YN$ and $YY$  interactions.
(iii) Using the interaction models in (i), accurate calculations of hypernuclear structure were performed. By comparing the theoretical results with experimental data, the interaction potentials were iteratively refined and improved.
Regarding the $\Lambda N$ interaction, structure calculations based on the Nijmegen models \cite{Nagels:1975fb, Nagels:1978sc, Maessen:1989sx, Rijken:1998yy} have been carried out within both the shell model \cite{Millener:2010zz} and the cluster model framework \cite{Hiyama:2013iya}. In parallel, a large number of high-resolution 
$\gamma$-ray spectroscopy experiments have been performed. By combining these experimental data with theoretical calculations, it has been possible to extract detailed information on the $\Lambda N$ interaction.

As a next step, it is crucial to obtain information on the strangeness 
$S=-2$ sector, particularly on the $\XiN$ interaction. 
In contrast to the 
$S=-1$ sector, the amount of experimental data available for 
$S=-2$ systems is still quite limited.
Recently, the bound state of $^{15}_\Xi{\rm C} (\Xi+^{14}$N)
 was observed for the first time \cite{Nakazawa:2015joa}.
 In 2024, new experimental data on the $^{12}{\rm C}(K^-,K^+) ^{12}_\Xi$Be 
 reaction were obtained at J-PARC \cite{Ichikawa:2024fjf}.
 In this experiment, two peaks were observed, and one of them corresponds to a 
$\Xi$-separation energy of $B_\Xi=8.9 \pm 1.4 {\rm \,(stat.)}^{+3.8}_{-3.1}$ (syst.)
MeV.
These two independent observations provide strong evidence that the 
$\Xi$-nucleus potential is attractive.

For further investigation of the $N\Xi$ interaction, it is essential to theoretically predict the level structure of novel $\Xi$ hypernuclei and to propose corresponding experiments using the $(K^-, K^+)$ reaction at J-PARC. To accomplish this, the following two requirements must be fulfilled: (1) the adoption of a reliable $N\Xi$ interaction, and (2) the execution of high-precision few-body calculations for $\Xi$ hypernuclei.

Regarding (1), 
it is most desirable to determine the $YN$ and $YY$ interactions from first-principles lattice QCD simulations. Recently, such simulations have been performed by the HAL QCD Collaboration using $(2+1)$-flavor lattice QCD on a large spacetime volume, $(8.1~\mathrm{fm})^4$, with nearly physical quark masses, $(m_\pi, m_K) = (146, 525)$ MeV.
Utilizing the HAL QCD method~\cite{Ishii:2006ec, Aoki:2009ji, Ishii:2012ssm},
various $YN$ and $YY$ interactions 
such as $\Lambda \Lambda$,
$N\Xi$, $N\Omega$ and $\Omega \Omega$
were determined~\cite{HALQCD:2019wsz, HALQCD:2018qyu, Gongyo:2017fjb, Aoki:2020bew}.
We note that the (spin-isospin-averaged) $N\Xi$ momentum correlation in $p$–$p$ collisions was investigated by the ALICE Collaboration at the CERN Large Hadron Collider (LHC), and the theoretical prediction based on the HAL QCD $N\Xi$ interaction was successfully confirmed by the experimental data~\cite{ALICE:2020mfd}. Under these circumstances, the study of $\Xi$ hypernuclei based on the HAL QCD $N\Xi$
 interaction is critically important for understanding 
$YN$ and $YY$
interactions.

Regarding (2), the Gaussian Expansion Method (GEM) is one of the most powerful methods for accurately describing both bound and resonant states in three- and four-body systems. GEM was originally proposed by Kamimura~\cite{Kameyama:1989zz} to perform three-body calculations of muonic molecules and muon-atomic collisions.
In a usual variational approach to three-body bound states, the Hamiltonian is diagonalized in a finite basis of square-integrable ($L^2$) functions. GEM employs a superposition of Gaussian basis functions with geometric progression ranges, allowing flexible and efficient representation of short- and long-range correlations. The method has been applied to the four-nucleon bound state problem using realistic $NN$ interactions~\cite{Kamada:2001tv}, and has achieved accuracy comparable to other sophisticated few-body methods such as the Faddeev approach.

By utilizing the $N\Xi$ interaction derived from first-principles lattice QCD calculations
with the HAL QCD method
and applying accurate few-body techniques such as GEM, a robust theoretical framework to study $YN$ and $YY$ interactions can be established. This strategic approach is illustrated schematically in Fig.~\ref{fig:strategy} (b).

In particular,
structure of the candidate of the lightest $\Xi$ hypernuclei such as $NN\Xi$
 and $NNN\Xi$ systems was studied and
energy spectra of $\XiNaa$ system were
discussed  \cite{Hiyama:2019kpw,Hiyama:2022jqh}.
In this paper, we review these studies.

\section{Method}\label{sec2}

\subsection{HAL QCD method}\label{hal}

\subsubsection{Basic formulation}

The HAL QCD method~\cite{Ishii:2006ec, Aoki:2009ji, Ishii:2012ssm, Aoki:2020bew, Aoki:2023qih} 
is a theoretical framework
which can determine hadron-hadron interactions
from first-principles lattice QCD simulations.
In this method, the key quantity is the equal time 
Nambu-Bethe-Salpeter (NBS) wave function, defined by
\begin{eqnarray}
\psi_{W}^{B_1B_2}({\bec{r}}) e^{-W t} \equiv {1\over \sqrt{Z_{B_1} Z_{B_2}}} \sum_{\bec{x}} \langle 0 \vert
B_1({\bec{x}}+{\bec{r}},t) B_2({\bec{x}}, t) \vert B_1B_2; W\rangle,
\end{eqnarray}
where
we consider a two-baryon system $B_1 B_2$ with single-baryon masses of $m_{B_1}$ and $m_{B_2}$
as a typical example.
$\vert B_1B_2; W\rangle$ is a QCD eigenstate 
with the center of mass (CM) energy 
$W= 
\sqrt{{\bec{p}}_W^2 + m_{B_1}^2} + 
\sqrt{{\bec{p}}_W^2 + m_{B_2}^2}
$ 
with ${\bec{p}}_W$ being a relative momentum,
and $Z_{B_{1,2}}$ are the wave function renormalization factors.
For simplicity, other quantum numbers such as spin and isospin are suppressed.
A baryon operator $B_{1,2}(x)$ with $x = (\bec{x},t)$ is  made of quarks,  
which is given by,
e.g., for local nucleon ($N$) and $\Xi$ operators,
\begin{eqnarray}
N_q(x) &=& \epsilon_{abc} \left( u^{T}_a(x) C\gamma_5 d_b(x) \right) q_c(x) \\
\Xi_q(x) &=& \epsilon_{abc} \left( s^{T}_a(x) C\gamma_5 q_b(x) \right) s_c(x), 
\end{eqnarray}
where $C=\gamma_4\gamma_2$ is a charge conjugation matrix, and
$N_{q=u (d)}$ for a proton (neutron)
and 
$\Xi_{q=u (d)}$ for a $\Xi^0 (\Xi^-)$ baryon.
For simplicity, we give formula in the case of
equal masses $m_{B_1} = m_{B_2} = m_B$ below.
Extensions to the case of unequal masses $m_{B_1}\not=m_{B_2}$ are straightforward.

If the total energy lies below the inelastic threshold as $W < W_{\rm th}$, 
the above NBS wave function satisfies the Helmholtz equation  at large $r\equiv\vert {\bec{r}}\vert>R$
as
\begin{eqnarray}
\left[ p_W^2 +\nabla^2\right] \psi_{W}^{B_1B_2}({\bec{r}}) \simeq 0, 
\quad  p_W \equiv\vert {\bec{p}}_W\vert,
\end{eqnarray}
where $R$  is a range of the interaction between two baryons,
and
the NBS wave function for given orbital angular momentum $\ell$ and total spin $s$ behaves asymptotically for large $r > R$
as
\begin{eqnarray}
 \psi_{W}^{B_1B_2}(r;\ell s)&\propto& {\sin ( p_W r -\ell\pi/2 +\delta_{\ell s}(W) ) \over p_W r},
\end{eqnarray}
where $\delta_{\ell s}(W)$ is a phase shift of the QCD S-matrix for $B_1 B_2$ scattering
at the CM energy $W$~\cite{Lin:2001ek,CP-PACS:2005gzm}.

Utilizing the fact that the NBS wave function contains the information
of the phase shift $\delta_{\ell s}(W)$, 
the HAL QCD method introduces a non-local but energy-independent potential $U({\bec{r}}, {\bec{r}}')$ from the NBS wave function as
\begin{eqnarray}
(E_W - H_0) \psi_{W}^{B_1B_2}({\bec{r}}) &=& \int d\bec{r}'\, U({\bec{r}}, {\bec{r}}')  \psi_{W}^{B_1B_2}({\bec{r}}'),
\quad E_W  = {p_W^2\over 2\mu}, \ H_0 = -{\nabla^2\over 2\mu}, 
\label{eq:nonlocal U}
\end{eqnarray}
where $\mu = m_B/2$ is a reduced mass.
By construction, 
it is guaranteed that, 
if one solves the Schr\"{o}dinger equation
with the potential $U({\bec{r}}, {\bec{r}}')$
in the infinite volume, scattering phase shifts $\delta_{\ell s}(W)$ for $W < W_{\rm th}$ are correctly reproduced.

In practical lattice QCD simulations,
direct determination of the non-local potential is difficult,
and it is convenient to introduce the velocity (derivative) expansion
for the non-locality of the potential,
$U({\bec{r}}, {\bec{r}}') = V({\bec{r}},{\bec{\nabla}}) \delta^{(3)}({\bec{r}} -{\bec{r}}')$, 
where
\begin{eqnarray}
\lefteqn{
 V({\bec{r}},{\bec{\nabla}}) =
 \underbrace{V_0(r) + V_\sigma(r) {\bec{\sigma}}_1\cdot  {\bec{\sigma}}_2 + V_{\rm T}(r) S_{12}}_{\rm LO}
} \nn
&& \qquad\qquad\qquad +
 \underbrace{V_{\rm SLS}(r){\bec{L}}\cdot{\bec{S}_+}
 + V_{\rm ALS}(r){\bec{L}}\cdot{\bec{S}_-}
 }_{\rm NLO} 
 + O(\nabla^2)
\end{eqnarray}
at lowest few orders~\cite{Okubo:1958qej}.
Here $V_0(r)$ is a central potential, $V_\sigma$ is a spin dependent central potential with a Pauli matrix $\bec{\sigma}_i$ acting on a spinor index of an $i$-th baryon, $V_{\rm T}(r)$ is a tensor potential with a tensor operator $S_{12}= 3({\bec{r}} \cdot{\bec{\sigma}}_1) ({\bec{r}} \cdot{\bec{\sigma}}_2)/r^2-  {\bec{\sigma}}_1\cdot  {\bec{\sigma}}_2$,  
and $V_{\rm SLS}(r)$  ($V_{\rm ALS}(r)$)
is a symmetric (antisymmetric) spin-orbit potential with an angular momentum ${\bec{L}} ={\bec{r}}\times {\bec{p}}$ and a spin operator
${\bec{S}_{\pm}}=({\bec{\sigma}}_1 \pm {\bec{\sigma}}_2)/2$.
Each coefficient function is further decomposed into its flavor components, e.g.,
$
V_0(r) = V_0^0(r) + V_0^\tau(r) {\bec{\tau}}_1\cdot{\bec{\tau}}_2
$
where ${\bec{\tau}}_i$ is a Pauli matrix acting on an isospin index of an $i$-th baryon.

Since NBS wave functions cannot be obtained directly in lattice QCD,
a correlation function for two baryons is considered instead,
\begin{eqnarray}
F_{J}^{B_1B_2}({\bec{r}}, t) = \sum_{\bec{x}} \langle 0 \vert B_1({\bec{x}}+{\bec{r}},t+t_0) B_2({\bec{x}},t + t_0) \bar{J}_{B_1B_2}(t_0)\vert 0\rangle,
\end{eqnarray}
 where $\bar{J}_{B_1B_2}(t_0)$ is a source operator which creates two-baryon states at time $t_0$.
 If a time separation $t$ is large enough to suppress inelastic contributions to the correlation function, it is shown that
 \begin{eqnarray}
F_{J}^{B_1B_2}({\bec{r}}, t) &=& \sum_n A_{J,n}^{B_1B_2}  \psi_{W_n}^{B_1B_2}({\bec{r}}) e^{-W_n t} +\cdots, 
\end{eqnarray}
where $A_{J,n}^{B_1B_2} =\sqrt{Z_{B_1} Z_{B_2}}\  \langle B_1B_2;W_n \vert \bar{J}_{B_1B_2} (0)\vert 0\rangle$ and ellipses represent contributions from inelastic states, 
which are suppressed as $e^{- W_{\rm th} t}$.
If we further take the limit of $t \to\infty$, the correlation function reduces to the NBS wave function for the ground state as
\begin{eqnarray}
\lim_{t \to\infty} F_{J}^{B_1B_2}({\bec{r}}, t) &=& A_{J,0}^{B_1B_2}   \psi_{W_0}^{B_1B_2}({\bec{r}}) e^{-W_0 t}
+ O\left( e^{-W_{n\not=0} t} \right),
\label{eq:groundF}
\end{eqnarray}
where $W_0$ is the lowest energy of $B_1B_2$ states.

The leading order potential, 
$
V^{\rm LO}({\bec{r}}) = 
V_0(r) +  V_\sigma(r) {\bec{\sigma}}_1\cdot  {\bec{\sigma}}_2 + V_{\rm T}(r) S_{12}, 
$
may be obtained from
$
\displaystyle
\lim_{t\to\infty} V^{\rm LO}({\bec{r}},t), 
$
where
\begin{eqnarray}
V^{\rm LO}({\bec{r}}, t) &=&
{ ( E_{W_0} -H_0)  F_J^{B_1B_2}({\bec{r}},t) \over  F_{J}^{B_1B_2}({\bec{r}}, t)} ,
\label{eq:pot_naive}
\end{eqnarray}
which is independent on the source operator$\bar{J}_{B_1B_2}$
as $t\to\infty$, 
since a source dependent constant $A_{J,0}^{B_1B_2}$ is canceled in the above ratio.

For the above procedure to work, the ground state saturation in $F_J^{B_1B_2}$, Eq.~(\ref{eq:groundF}), must be satisfied by taking a sufficiently large $t$
compared to the (inverse of) the energy gap between
the ground state and the excited states, $(W_1 - W_0)^{-1}$.
Since a typical energy gap is small which scales as 
$W_1 - W_0 = O( L^{-2} )$ in lattice QCD on a finite box with a spacial volume $L^3$,
it is necessary to take a large $t$,
which however is impractical since 
the statistical fluctuations of 
$F_J^{B_1B_2}$ 
grows exponentially with a larger $t$, 
known as a signal-to-noise problem~\cite{Parisi:1983ae, Lepage:1989hd}.
In order to overcome this difficulty, 
an alternative extraction of potentials,
called "time-dependent HAL QCD method", has been introduced~\cite{Ishii:2012ssm} as will be explained below.

\subsubsection{Time-dependent HAL QCD method}

For a correlation function in a usual lattice QCD calculation,
the contribution from the state of the interest (e.g., the ground state)
serve as the signal, while 
those from other states (e.g., excited states) give noises (i.e., systematic errors).
This is why one has to disentangle contributions from each state 
by taking, e.g., $t\to\infty$, but then one encounters the signal-to-noise problem.

The key advantage of the time-dependent HAL QCD method is that
one can determine the potential $U({\bec{r}}, {\bec{r}}')$
directly from a correlation function without disentangling the contributions from each state,
by noting that the potential is energy-independent
and thus one can extract the signal of the potential from 
not only the ground state but also (elastic) excited states which lie below $W_{\rm th}$.

To be more specific, 
the following ratio of correlation functions is considered in the time-dependent HAL QCD method,
\begin{eqnarray}
R_J^{B_1B_2}({\bec{r}},t) = {F_J^{B_1B_2}({\bec{r}},t) \over G_{B_1}(t) G_{B_2}(t)}, \quad
G_B(t) =\sum_{\bec{x}} \langle 0\vert B({\bec{x}},t) \bar B({\bec{0}},0)\vert 0 \rangle \simeq Z_B e^{-m_B t} + \cdots, ~~~~~
\end{eqnarray}
which behaves
\begin{eqnarray}
R_J^{B_1B_2}({\bec{r}},t) &\simeq&\sum_n {A_{J,n}^{B_1B_2}\over Z_{B_1} Z_{B_2}}  \psi_{W_n}^{B_1B_2}({\bec{r}}) e^{-\Delta W_n t},
\quad \Delta W_n = W_n - 2 m_B,
\end{eqnarray}
for $ t \gg 1/W_{\rm th}$, where inelastic contributions can be neglected. Using
$
\Delta W ={p_{W}}^2/m_B -{(\Delta W)^2} / (4 m_B),
$
one can derive the master equation for the time-dependent HAL method~\cite{Ishii:2012ssm},
\begin{eqnarray}
\left\{ - H_0 -{\partial \over \partial t} +{1\over 4 m_B} {\partial^2\over \partial t^2} \right\}
R_J^{B_1B_2}({\bec{r}},t) =  
\int d\bec{r}'\, U({\bec{r}}, {\bec{r}}')  R_J^{B_1B_2}({\bec{r}'},t) ,
\label{eq:t-dep}
\end{eqnarray}
where one may in practice determine the potential with the velocity expansion as mentioned before.
Note that 
the required condition, 
$t \gg 1/W_{\rm th}$ (saturation of elastic states),
becomes much easier to be achieved than the usual condition, 
$ t \gg 1/(W_1-W_0) \simeq O(L^2)$ (saturation of the ground state).
In fact, it was explicitly shown that the above feature is 
essential to perform reliable lattice calculations 
for two-baryon systems~\cite{Iritani:2018vfn, BaSc:2025yhy}.

\subsubsection{HAL QCD method for coupled channel systems}

It is possible to extend the HAL QCD method to coupled channel systems~\cite{Aoki:2011gt, Aoki:2012bb},
such as $\Lambda\Lambda$-$N\Xi$ systems.
In the following, we consider a case that $X_1+X_2\to Y_1+Y_2$ scattering occurs 
in addition to an elastic scattering $X_1+X_2\to X_1+X_2$, 
where $X_{1,2},Y_{1,2}$ are single-hadrons, and $  m_{X_1}+m_{X_2}  <  m_{Y_1} + m_{Y_2} < W < W_{\rm th}$ with $W$ and $W_{\rm th}$ being the total energy and the inelastic threshold of the coupled channel system, respectively.  For simplicity, $ m_{X_1}=m_{X_2}=m_X$ and
 $m_{Y_1}=m_{Y_2}=m_Y$ are assumed in the following.

NBS wave functions are generalized as
\begin{eqnarray}
\Psi^{X}_W ({\bec{r}} )e^{-W t}  &=& \frac{1}{\sqrt{Z_{X_1} Z_{X_2}}} \sum_{\bec{x}} \langle 0 \vert X_1({\bec{x}} +{\bec{r}},t) X_2 ({\bec{x}},t)\vert X+Y; W \rangle ,  \\
\Psi^{Y}_W ({\bec{r}} )e^{-W t}  &=& \frac{1}{\sqrt{Z_{Y_1} Z_{Y_2}}} \sum_{\bec{x}} \langle 0 \vert Y_1({\bec{x}} +{\bec{r}},t) Y_2 ({\bec{x}},t)\vert X+Y; W \rangle ,
\end{eqnarray}
where $\vert X+Y; W\rangle$ is a QCD eigenstate in the coupled channel system, which may be expressed as
\begin{eqnarray}
\vert X+Y;W\rangle &=& c_{X} \vert X_1X_2;W\rangle + c_Y  \vert Y_1Y_2;W\rangle +\cdots, \\
\vert H_1H_2;W\rangle &=&  \sum_{\vert{\bec{p}}\vert = p_W^H} \vert H_1, {\bec{p}} \rangle_{\rm in} \otimes   \vert H_2, -{\bec{p}} \rangle_{\rm in}
\end{eqnarray}
for $H=X,Y$. 
Here $\vert H_{1,2}, {\bec{p}} \rangle_{\rm in}$ is an in-state for a hadron $H_{1,2}$ with a momentum ${\bec{p}}$, and  
$p_W^H$ is given by $W=2\sqrt{(p_W^H)^2+m_H^2}$ for the channel $H=X,Y$.

A coupled channel non-local and energy-independent potential is defined  by
\begin{eqnarray}
\left[ {(p_W^H)^2\over 2\mu_H} +{\nabla^2\over 2\mu_H} \right] \Psi_W^H({\bec{r}}) = \sum_{H'=X,Y}\int d\bec{r}'\, U^H{}_{H'}({\bec{r}},{\bec{r}'} )  \Psi_W^{H'}({\bec{r}'}), 
\end{eqnarray}
where $\mu_H= m_H/2$ is  a reduced mass in a channel $H$.

As in the case of the single channel,
it is better to use time-dependent formalism for the coupled channel case.
We consider the correlation function defined by
\begin{eqnarray}
F_J^H({\bec{r}}, t) &=& \sum_{\bec{x}} \langle 0\vert H_1({\bec{x}}+{\bec{r}},t) H_2({\bec{x}},t) \bar{J}_H (0) \vert 0\rangle \\
R_{J_i}^H({\bec{r}}, t) &=& {F_{J_i}^H({\bec{r}},t)\over G_{H_1}(t) G_{H_2}(t)}
\simeq \sum_n  {A_{J_i,n}^H\over Z_{H_1} Z_{H_2}} \Psi_{W_n}^H({\bec{r}}) e^{- \Delta W_n^H t}, 
\end{eqnarray}
for $H=X,Y$ and $ i=1,2$,
where $J_i$ represent two different source operators, $\Delta W_n^H=W_n - m_{H_1}-m_{H_2}$, and 
$t$ is taken to be large enough to ignore inelastic contributions.
One can then show the following time-dependent HAL QCD formula for a coupled channel system,
\begin{eqnarray}
\left( {\nabla^2\over 2\mu_H}-{\partial\over \partial t} + {1\over 8\mu_H} {\partial^2\over \partial t^2} \right) R_{J_i}^H({\bec{r}}, t) 
&\simeq & \sum_{H'} \Delta^{H}{}_{H'}(t) \int d\bec{r}' \ U^H{}_{H'}({\bec{r}}, {\bec{r}'}) R_{J_i}^{H'}({\bec{r}'}, t),~~~~
\label{eq:t-dep_C}
\end{eqnarray}
where
\begin{eqnarray}
 \Delta^{H}{}_{H'} (t) = \sqrt{ Z_{H'_1} Z_{H'_2}\over Z_{H_1} Z_{H_2}} { e^{ - (m_{H'_1} + m_{H'_2}) t}\over e^{ - (m_{H_1} + m_{H_2}) t}}
\end{eqnarray}
is a factor which accounts for differences in masses and $Z$ factors between two channels.

\subsection{Gaussian Expansion Method}\label{subsec2}

To solve three- and four-body systems, one of the most powerful methods is the Gaussian Expansion Method (GEM), originally proposed by Kamimura in 1989~\cite{Kameyama:1989zz}.
This method was first applied to high-precision calculations of muonic molecular and muon-atomic collisions.
Since then, it has been extensively applied to a wide range of three-, four-, and five-body systems \cite{Hiyama:2003cu}.
For simplicity, we illustrate the method for a spinless three-body system.
To accurately describe such a system, three sets of Jacobi coordinates are introduced, as shown in Fig.~2.
The Hamiltonian of the system is described as
\begin{figure*}[htb]
\begin{center}
\includegraphics[scale=0.45]{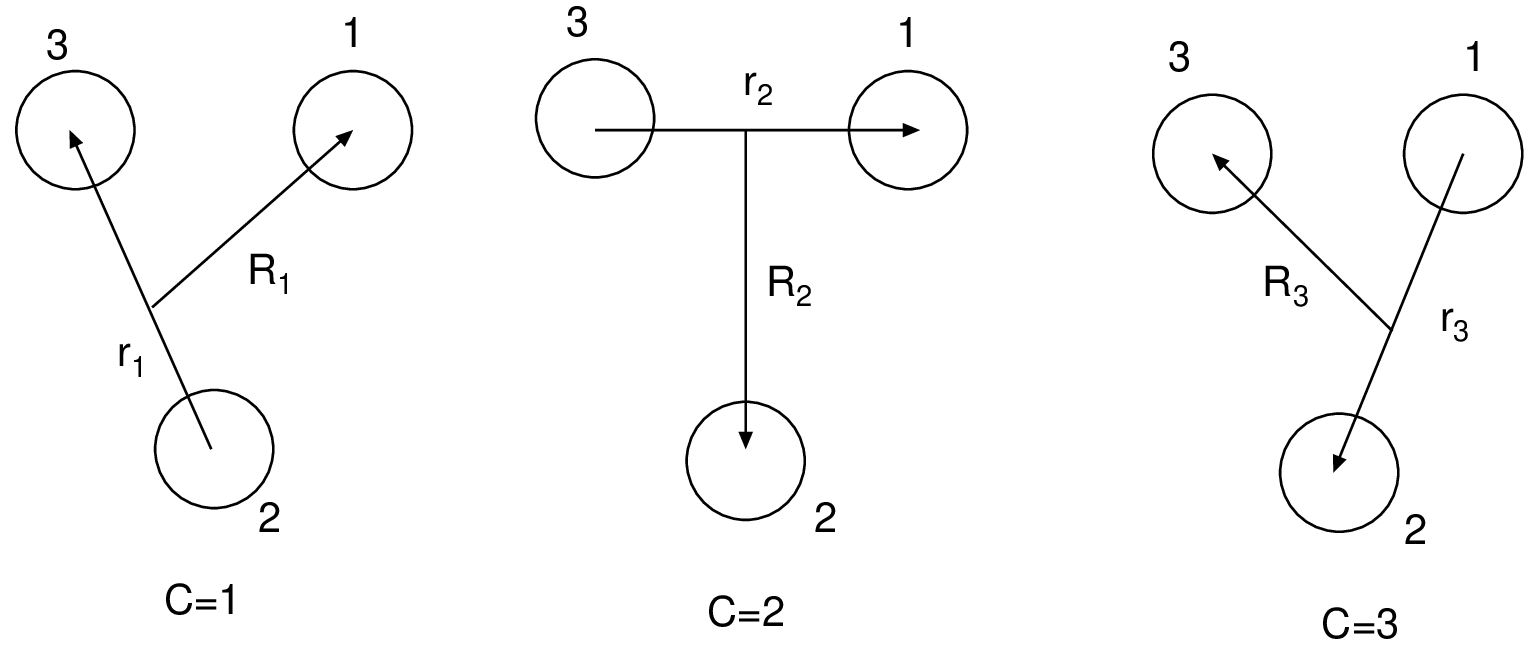}
\end{center}
\caption{Jacobi coordinate of three-body system.}
\label{fig:fig1}
\end{figure*}

\begin{equation}
H=-\frac{\hbar^2}{2\mu_{r_c}}\nabla_{r_c}^2-\frac{\hbar^2}{2\mu_{R_c}}\nabla_{R_c}^2
+V(r_1)+V(r_2)+V(r_3)
\end{equation}
where $c=1,2$ or 3 and $\mu_{r_c}=1/2m$ and $\mu_{R_c}=2/3m$.
Here $m$ is the mass of a particle in the three-body system.
In order to solve the Schr\"{o}dinger equation with this Hamiltonian,
the wavefunction is expanded in terms of the following
three-body basis functions:
\begin{equation}
\Psi_{JM} =\Phi_{JM}({\bf r}_1, {\bf R}_1) +\Phi_{JM}({\bf r}_2, {\bf R}_2)
+\Phi_{JM}({\bf r}_3, {\bf R}_3).
\end{equation}
These basis functions, $\Phi_{JM}({\bf r}_c, {\bf R}_c)$,
which are nonorthogonal to each other, are constructed on the
full set of three Jacobi coordinate systems.
Each basis function is expressed as
\begin{equation}
\Phi_{JM}({\bf r}_c, {\bf R}_c)=A^{c}_\alpha \phi_{n\ell}(r_c)\phi_{NL}(R_c)[Y_\ell({\bf \hat{r_c}})
Y_L({\bf \hat{R_c}})]_{JM} ,
\end{equation}
where
$\phi_{n \ell}(r)$ and $\phi_{NL}(R)$
are given by
\begin{eqnarray}
\label{eq:range}
\phi_{n \ell}(r)= r^\ell e^{-(r/r_n)^2} ,  \quad  r_n=r_1a^{n-1} \enskip (n=1, \cdot \cdot
, n_{\rm max})  \\ \nonumber
\phi_{NL}(R)=R^L e^{-(R/R_N)^2}, \quad R_N=R_1 A^{N-1} \enskip
(N=1, \cdot \cdot , N_{\rm MAX}).
\end{eqnarray}

The $[Y_\ell({\bf \hat{r_c}})
Y_L({\bf \hat{R_c}})]_{JM}$ is defined by
\begin{equation}
[Y_\ell({\bf \hat{r_c}})
Y_L({\bf \hat{R_c}})]_{JM}=\sum_{m M}(\ell m L M|JM)
Y_{\ell m}({\bf \hat{r_c}})
Y_{LM}({\bf \hat{R_c}})
\end{equation}

The eigen energies $E$ and amplitudes $A^{c}_\alpha$ are determined by the Ray-Ritz variational principle.
By solving the eigenvalue problem, we obtain not only the lowest state but also
the excited-state eigen functions with the same $J$ and parity.

An important issue in the variational method is how to select
a good set of basis functions.
The advantage of using a Gaussian function for radial part is
that calculations of matrix elements of the Hamiltonian becomes easier.
Also, by taking a geometric progression for the radial part  in Eq.~(\ref{eq:range}),
we can simultaneously describe proper short-range correlations
and the asymptotic behaviour of the total wave function.

In order to show how accurately our method provides binding energies and 
the wavefunction,
in Ref.~\cite{Kamada:2001tv}, we performed a benchmark test calculation 
of a four-nucleon
system which is useful for testing the method and the scheme 
of the calculation.
The benchmark calculation was performed among seven  methods:
the Faddeev-Yakubovsky equation method (FY), 
the Gaussian expansion method (GEM),
the Stochastic variation method (SVM),
the Hyperspherical Harmonic Variational Method (HH),
the Green's Function Monte Carlo (GFMC) method, the No-Core
Shell Model (NCSM), and the Effective Interaction
Hyperspherical Harmonic Method (EIHH).
The calculated binding energies and $r.m.s$ radii were compared 
and the values among the seven methods were
 in good agreement within three digits or within 
0.5 \%.

The GEM method has been mainly applied to hypernuclear structure.
As mentioned in the Introduction,
one of the primary goals of hypernuclear physics is to extract information on 
baryon-baryon interaction.
However, due to the lack of $YN/YY$ scattering experiments
  (only a few $YN$ and no $YY$ scattering data are available),
there are large uncertainties in the $YN/YY$ potentials.
As an alternative procedure, accurate calculation of hypernuclei using 
the shell model and
few-body method such as GEM have been performed, and
important constraints for the $YN$  interaction were given, especially for
the $\Lambda N$ interaction \cite{Hiyama:2013iya}.

As a next step, it is important to study the $S=-2$ sector such as the
 $\Lambda \Lambda$
interaction and the $N\Xi$ interaction.
Regarding the $\Lambda \Lambda$ interaction, we could extract information from
the binding energies of several observed double $\Lambda$ hypernuclei such as
$^6_{\Lambda \Lambda}$He\cite{Takahashi:2001nm}, $^{10}_{\Lambda \Lambda}$Be \cite{Nakazawa:2010zzb}
and $^{13}_{\Lambda \Lambda}$B \cite{Aoki:1991ip}.
Furthermore, femtoscopic analyses of  two-particle correlations in
high-energy $pp$, $pA$ and $AA$ collisions at RHIC \cite{STAR:2014dcy}
and LHC \cite{ALICE:2018ysd} have started to provide information of the
 $\Lambda \Lambda$
interaction.
 Regarding the $N\Xi$ interaction, through the observation of  a bound 
$\Xi$ hypernucleus, $^{15}_\Xi$C ($^{14}{\rm N} +\Xi)$,
we found that the  $\Xi$-nucleus interaction should be attractive.

\begin{figure*}[htb]
\begin{center}
\begin{minipage}{0.45\hsize}
\includegraphics[scale=0.40]{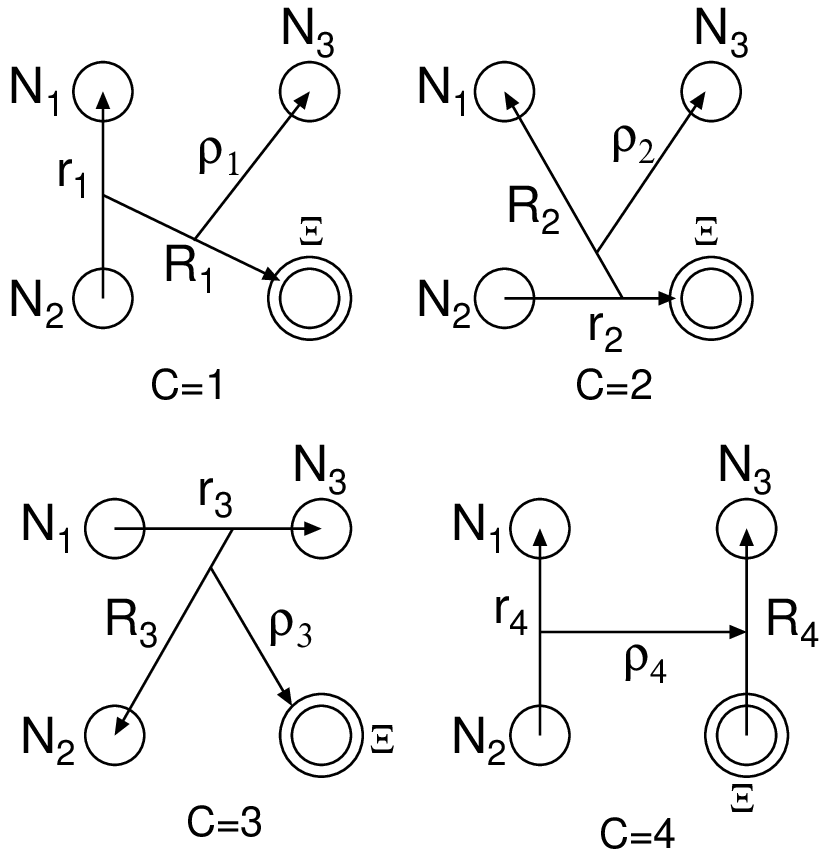}
\includegraphics[scale=0.40]{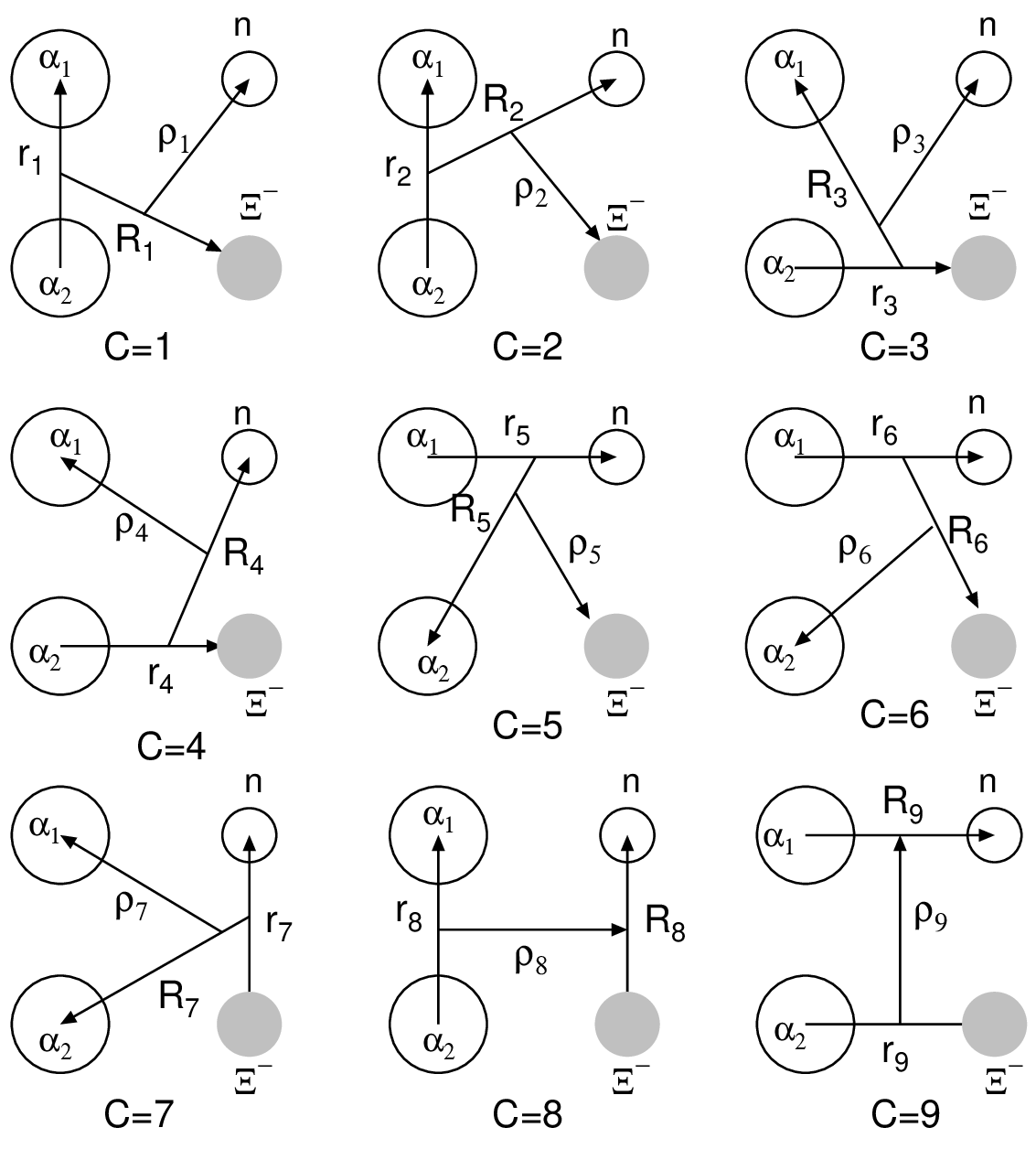}
\end{minipage}
\end{center}
\caption{Jacobi coordinate of $NNN\Xi$ and
$\XiNaa$ systems.}
\label{fig:fig2}
\end{figure*}

Motivated by these observations,
it is important to study what kinds of
$\Xi$-hypernuclei should be observed as bound states,
and
how each  spin-isospin component of  the $N\Xi$ interaction
contributes to the binding energies.
For this purpose,
we study the $NN\Xi$, $NNN\Xi$ and $\XiNaa$ systems~\cite{Hiyama:2019kpw,Hiyama:2022jqh}.
Employing the $N\Xi$ interaction
obtained from first-principles lattice QCD simulations using  the HAL QCD method,
we solve the Schr\"{o}dinger equation using the  GEM,
where the Jacobi coordinates shown in Fig.~\ref{fig:fig1} are used for the $NN\Xi$,
and those in Fig.~\ref{fig:fig2} are used for the $NNN\Xi$ and $\XiNaa$.

\section{Results}\label{results}

\subsection{Lattice QCD results for the $N\Xi$ interactions}\label{lat-NX}

The $N\Xi$ potentials are obtained from first-principles (2+1)-flavor lattice QCD simulations
using the (time-dependent) HAL QCD method.
The lattice simulations are performed
with the Iwasaki gauge action and the non-perturbatively ${\cal O}(a)$-improved Wilson quark action
together with the stout smearing,
   in a large spacetime volume, $L^4 = (8.1 \ {\rm fm})^4$ ($(L/a)^4=96^4$),  with nearly physical quark masses,
   $(m_{\pi}, m_K)$=(146, 525) MeV, at a lattice spacing, $a = 0.0846$ fm.  
A series of simulations with the same setup
successfully determined not only the $\XiN$ interactions~\cite{HALQCD:2019wsz}
but also 
$N\Omega$~\cite{HALQCD:2018qyu}, 
$\Omega\Omega$~\cite{Gongyo:2017fjb},
$\Omega_{ccc}\Omega_{ccc}$~\cite{Lyu:2021qsh},
$N\phi$~\cite{Lyu:2022imf},
$DD^*$~\cite{Lyu:2023xro} and
$NJ/\psi, N\eta_c$~\cite{Lyu:2024ttm} interactions.

The $\XiN$ system can couple to 
$\Lambda\Lambda$, $\Lambda\Sigma$ and $\Sigma\Sigma$ systems if one considers 
the two octet baryon degrees of freedom
(Note, however, that some of these couplings are forbidden
depending on the quantum numbers of the system.)
Since we consider $\Xi$-hypernuclei at low energies in this paper,
we determine the $\XiN$ effective central potentials 
considering only the two lowest channels, $\Lambda\Lambda$ and $\XiN$.
Such potentials are still relevant for the energy region 
below the $\Lambda\Sigma$ threshold, where the effect from higher channels,
$\Lambda\Sigma$, $\Sigma\Sigma$, are effectively included.
Since the $\Lambda\Lambda$ can couple to $\XiN$ only in the $^{11}$S$_0$ channel,%
\footnote{
Here,    
we employ the spectroscopic notation $^{2T+1,2s+1}{\rm S}_J$
to classify the S-wave $\XiN$  interaction, where $T$, $s$, and $J$ stand for total isospin, total spin, and total angular momentum, respectively.
}
we perform the 
$2\times 2$ coupled channel analysis for the $^{11}$S$_0$ channel,
whereas the $\XiN$ single channel analysis is performed for other channels,
$^{31}$S$_0$, 
$^{13}$S$_1$  and
$^{33}$S$_1$.

We calculate the effective central potentials at the leading order in the derivative expansion
at the imaginary-time distances $t/a=11,12,13$.
Ideally, the result is independent of $t/a$, while
there exists $t/a$-dependence in the actual result,
and therefore $t/a$-dependence is used to estimate systematic errors in lattice calculations.
The obtained potentials together with corresponding phase shifts
as well as 
further details of the lattice calculations 
are given in Ref.~\cite{HALQCD:2019wsz}.
Note that 
while 
the $\Lambda\Lambda$-$\XiN$ coupled channel analysis
in the $^{11}{\rm S}_0$ channel shows that
the $\XiN$ interaction in this channel is moderately attractive,
deeply bound $H$-dibaryon is not found below the $\Lambda\Lambda$ threshold.
We also note that one can make a prediction on $p$-$\Xi^-$ correlation function
in ultrarelativistic nucleus-nucleus collision (so-called femtoscopic study) based on 
the HAL QCD $\XiN$ potentials,
and such theoretical prediction was confirmed by 
$pp$ collision experiment at the CERN LHC
by the ALICE Collaboration~\cite{ALICE:2020mfd, Kamiya:2021hdb}.

  To make the few-body calculation feasible, we fit  the lattice QCD result of the 
 potentials
    with multiple  Gaussian forms at short distances and the Yukawa  form  
    with  form factors at medium to long distances~\cite{HALQCD:2019wsz}.  As for the pion and Kaon masses which 
     dictate the long range part of the potential, we use 
     $(m_{\pi}, m_K) = (146, 525)$ MeV to fit the 
      lattice data, and take 
      $(m_{\pi}, m_K) = (138, 496)$ MeV 
      for calculating the $\Xi$-nucleus systems~\cite{Hiyama:2019kpw, Hiyama:2022jqh, Kamiya:2021hdb}.
   In the $^{11}{\rm S}_0$ channel, the channel-coupling
      between  $\Lambda\Lambda$ and $\XiN$ is found to be  weak~\cite{HALQCD:2019wsz}.
We therefore introduce an effective single-channel $\XiN$ potential in which 
   the coupling to $\Lambda\Lambda$  in $^{11}${\rm S}$_0$  is renormalized into  
   a single range Gaussian form   
$U_2 \cdot {\rm exp}(-(r/\gamma)^2)$ with $\gamma$=1.0 fm with $U_2 (<0)$ chosen to reproduce the 
 $\XiN$ phase shifts obtained with channel coupling.

\begin{figure*}[htb]
\begin{center}
\begin{minipage}{0.45\hsize}
\includegraphics[scale=0.4]{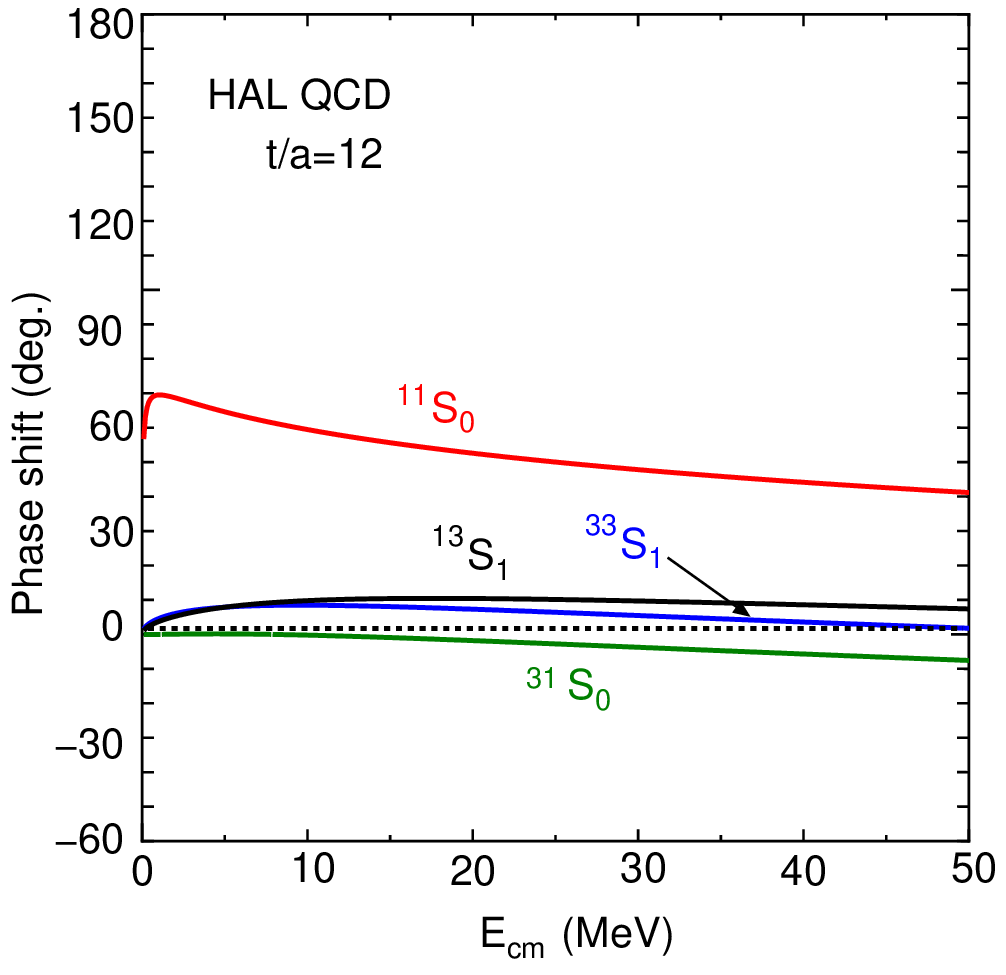}
\end{minipage}
\end{center}
\caption{$\XiN$ phase shifts in the 
$^{11}{\rm S}_0$, 
$^{13}{\rm S}_1$,
$^{33}{\rm S}_1$ and 
$^{31}{\rm S}_0$ 
channels using
the HAL QCD potential ($t/a=12$)~\cite{Hiyama:2019kpw}.
}
\label{fig:phase}
\end{figure*}

In Fig.~\ref{fig:phase}, we show  the $\XiN$ phase shifts calculated with  
the HAL QCD method at $t/a=12$.
The  statistical and systematic errors 
  are not shown in Fig.~\ref{fig:phase},  
  but are taken into account in the few body calculations below.
   From the figure, one finds that
   the $\XiN$ interaction in the $^{11}${\rm S}$_0$ channel is moderately attractive,
   while those in other channels are much weaker:
   the $^{13}$S$_1$ and $^{33}$S$_1$ channels are weakly attractive and
  the $^{31}$S$_0$ channel is weakly repulsive.

\subsection{Level structure of $\XiNN$ and $\XiNNN$ systems}\label{lat-NNNX}

For the $\XiNN$ system, we have calculated the binding energies in the $(T,J^\pi) = (1/2,3/2^+)$ and $(1/2,1/2^+)$ states
 using the HAL QCD $\XiN$ potential at time slices $t/a = 11$, 12, and 13 together with AV8 $NN$ potentials.
 However, no bound states are found for either state  at any of these time slices.

In the case of the $\XiNNN$ system, we investigate the binding energies in
 the $(T,J^\pi) = (0,0^+)$, $(0,1^+)$, $(1,0^+)$, and $(1,1^+)$ states.
 Among them,  only the $(0,1^+)$ state exhibits a weakly bound configuration. 
The corresponding binding energies with respect to the ${}^3\mathrm{H}/{}^3\mathrm{He}+\Xi$ threshold are 0.63 MeV, 0.36 MeV, and 0.18 MeV for $t/a = 11$, 12, and 13, respectively. No bound states are observed in  other channels.

The emergence of a bound state in the $(0,1^+)$ channel can be attributed to the partial-wave contributions of the $\XiN$ interaction. 
The average  $\XiN$ potential, $\bar{V}_{\XiN}$,
can be expressed as
$$
\bar{V}_{N \Xi}=
\frac{1}{6}V_{N \Xi}(\mbox{$^{11}$S$_0$}) +
\frac{1}{3}V_{N \Xi}(\mbox{$^{13}$S$_1$}) + 
\frac{1}{2}V_{N \Xi}(\mbox{$^{33}$S$_1$})
$$
for the $1^+$ state.
 According to the HAL QCD potential, the ${}^{11}$S$_0$ channel exhibits the strongest attraction, while the ${}^{13}$S$_1$ and ${}^{33}$S$_1$ channels are also weakly attractive. Thus, in the $(0,1^+)$ state, all contributing components are attractive, leading to a weakly bound state, particularly due to the strong attraction in the ${}^{11}$S$_0$ channel.
In contrast, the other spin-isospin configurations either lack the ${}^{11}$S$_0$
 component or involve only a small contribution from it, which explains the absence of bound states in those channels.

Here, it should be noted that the $\XiNNN$ state is above the $NN\Lambda \Lambda$
threshold. Then, this state should decay into $\Lambda \Lambda$ state by strong
interaction, $\XiN-\Lambda \Lambda$ coupling.
We estimate the decay width $\Gamma$ perturbatively using the  $\XiN-\Lambda \Lambda$ coupling
potential in the HAL QCD potential, and found that $\Gamma = $
0.06, 0.05, 0.03 MeV
at $t/a=11,12,13$, respectively.
We see that the decay widths are quite small.
One possible way to produce this state is through heavy-ion collisions at GSI and CERN LHC.

 \subsection{Level structure of $\XiNaa$ system}  \label{lat-NX1}

As discussed in the previous subsection, the contribution of specific spin-isospin components of the $\XiN$ interaction plays a crucial role in forming bound states in the $\XiNNN$ system.
It is important to confirm the spin-isospin dependence of the
HAL QCD $\XiN$ interaction experimentally.
Ideally, this can be done by $\XiN$ scattering experiments.
However, such measurements are extremely challenging due to the short lifetime and limited availability of $\Xi$ beams.

As an alternative approach, the study of the structure of $\Xi$ hypernuclei provides valuable information. In particular, the $\XiNa$ and $\XiNaa$ systems are well suited for probing the spin-isospin components of the $\XiN$ interaction. This is because the 
$\alpha$ particle is a spin-isospin saturated system, and thus the level structures of these few-body systems are directly linked to the properties of the $\XiN$ two-body interaction.

In this work, we calculate the energy spectra of the $\XiNa$ and $\XiNaa$ systems to extract insights into the spin-isospin dependence of the $\XiN$ interaction.
In the case of the $\XiNa$ system, the total isospin can be either $T = 1$ or $T = 0$. Since the ground state of $^5$He has $J^\pi = 3/2^-$ (in the simple shell-model configuration, the two protons and two neutrons occupy the $0s_{1/2}$ orbit, while the valence neutron occupies the $0p_{3/2}$ orbit),
the total spin-parity and isospin configurations of the $\XiNa$ system are $(T, J^\pi) = (1, 1^-)$, $(1, 2^-)$, $(0, 1^-)$, and $(0, 2^-)$. Among these, the $J^\pi = 2^-$ states correspond to spin-parallel $\XiN$ pairs, while the $1^-$ states correspond to spin-antiparallel pairs.
A similar classification applies to the $\XiNaa$ system. Since the core nucleus $^9$Be has $J^\pi = 3/2^-$, the total isospin-spin configurations of the $\XiNaa$ system are also $(T, J^\pi) = (1, 1^-)$, $(1, 2^-)$, $(0, 1^-)$, and $(0, 2^-)$.
The spin-isospin channels of the $\XiN$ pair
are related to these four configurations, namely,  $(1,1^-)$, $(1,2^-)$, $(0,1^-)$, and $(0,2^-)$ are dominated by
${}^{31}$S$_0$, ${}^{33}$S$_1$, ${}^{11}$S$_0$, and ${}^{13}$S$_1$ components, respectively.

For the $\XiNa$ system, our calculations show that no bound states appear in the $T = 1$ sector for either the $J^\pi = 1^-$ or $2^-$ states. In the $T = 0$ sector, although the 
${}^{11}$S$_0$ channel is the most attractive among the HAL QCD $\XiN$ potentials, its attraction is not sufficient to form a bound state in the $J^\pi = 1^-$ channel. However, we note that a very weakly bound state may appear in this channel when using the HAL QCD potential at $t/a = 11$.
Experimentally, the $\XiNa$ system with total isospin $T = 1$ and $T = 0$ could potentially be produced via the $(K^-, K^+)$ and $(K^-, K^0)$ reactions on a $^6$Li target, respectively. Nevertheless, the existence of bound states in $A = 6$ hypernuclear systems is unlikely based on the HAL QCD $\XiN$ interaction.

\begin{figure*}[htb]
\begin{center}
\begin{minipage}{0.45\hsize}
\includegraphics[scale=0.45]{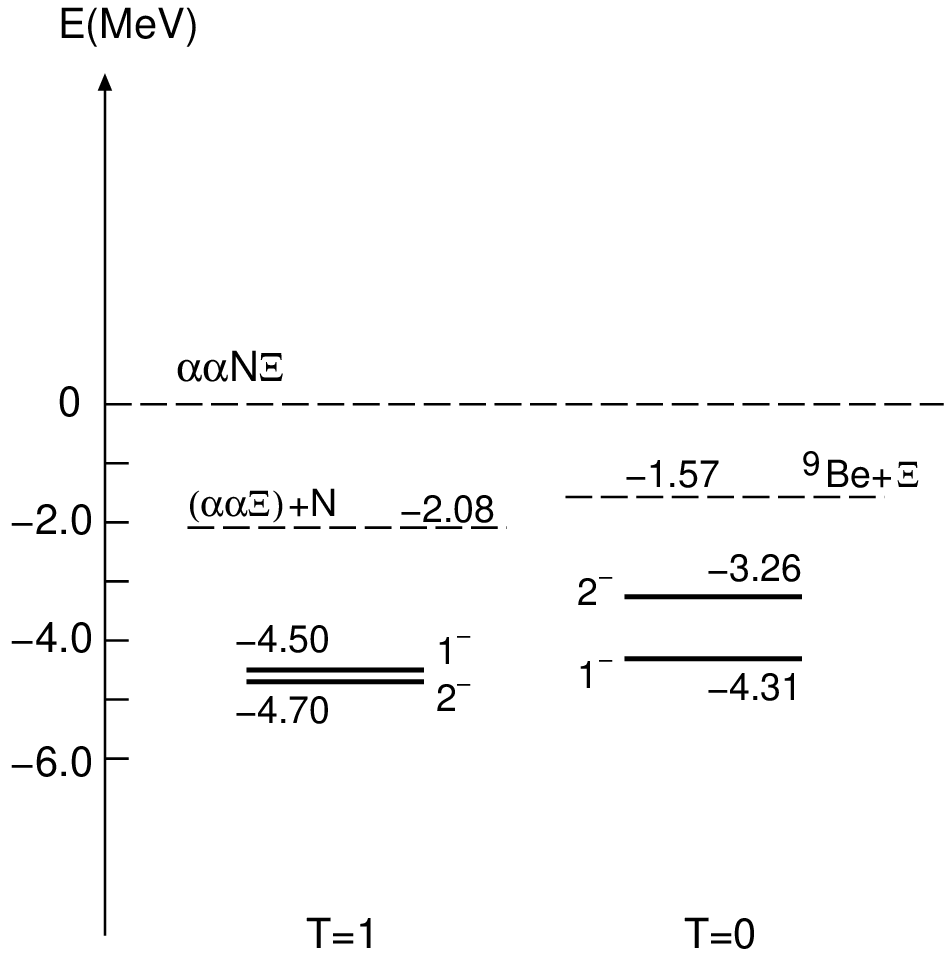}
\end{minipage}
\end{center}
\caption{The energy levels of $\alpha \alpha N \Xi$ system for $T=0$ and $T=1$~\cite{Hiyama:2022jqh}.
The energies are measured with respect to  $\alpha \alpha N \Xi$ breakup threshold.}
\label{fig:spectra}
\end{figure*}

We next consider the addition of another $\alpha$ particle to the $\XiNa$ system, resulting in the four-body $\XiNaa$ system for 
$T=1$ and 
for $T=0$, which is expected to be bound 
as suggested in Ref.~\cite{Hiyama:2008fq}.
The binding energy of the core nucleus $^9$Be ($T = 1/2$, $J^\pi = 3/2^-$) is 1.57 MeV. As a relevant subsystem, we also evaluate the $\XiMaa$ system in the $\XiNaa$ configuration with $(T, J^\pi) = (1/2, 1/2^+)$, obtaining a binding energy of $2.08^{+0.77}_{-0.63}$ MeV.
The calculated energy levels of the $\alpha \alpha N \Xi$ system for $(T, J^\pi) = (1, 1^-)$, $(1, 2^-)$, $(0, 1^-)$, and $(0, 2^-)$ are shown in Fig. \ref{fig:spectra}
together with Table I, 
where the central values and statistical errors are evaluated by the HAL QCD potential at $t/a = 12$,
  and the systematic errors are estimated by the potential at $t/a = 11$ and $13$.
The $J^\pi = 1^-$ and $2^-$ states in the $T = 1$ sector are bound with respect to the $(\Xiaa) + N$ breakup threshold. In the $T = 0$ sector, both states are bound with respect to the $^9$Be$ + \Xi$ threshold.

For $T = 1$, the $2^-$ state is the ground state due to the attractive nature of the ${}^{33}$S$_1$ component
 and the repulsive nature of the ${}^{31}$S$_0$ component. 
Since both the attraction and repulsion are relatively weak, the energy splitting between the $1^-$ and $2^-$ states is small. On the other hand, for $T = 0$, the $J^\pi = 1^-$ state is the ground state, mainly due to the moderate attraction in the 
${}^{11}$S$_0$ component.
On the other hand, the $^{13}$S$_1$ component
 is only weakly attractive, resulting in the $2^-$ state lying above the $1^-$ state, and
 the energy splitting between these spin-doublet states becomes larger than that in the $T=1$. It is noteworthy that the binding energy of the $(T=0, J^\pi = 1^-)$ state is the largest relative to the two-body breakup threshold ($^9\text{Be}+\Xi$ for $T=0$ and $(\Xiaa) + N$ for $T=1$) as can be seen in Fig.~\ref{fig:spectra}.
This is due to the strongest attraction in the $^{11}$S$_0$
component 
compared to other components.

\begin{table}[h]
\caption{Calculated energy levels and decay widths of
$\alpha \alpha N \Xi$ systems for $(T,J^{\pi})=(1,1^-),
(1,2^-), (0,1^-)$ and $(0,2^-)$.
The values are taken from Ref.~\cite{Hiyama:2022jqh}.
The energies are measured with respect to the $\XiNaa$ four-body breakup threshold.}\label{tab1}%
\begin{tabular}{@{}llll@{}}
\toprule
 & $J^\pi$  & $1^-$ & $2^-$\\
\midrule
$T=1$    & $E$(MeV)   & $-4.50^{+1.04}_{-0.80}$
& $-4.70^{+1.09}_{-0.83}$  \\
    & $\Gamma$(MeV)   & $0.02^{+0.01}_{-0.01}$
    & $0.02^{+0.01}_{-0.00}$  \\
$T=0$    & $E$(MeV)   & $-4.31^{+1.28}_{-1.04}$
  & $-3.26^{+1.10}_{-0.90}$  \\
    & $\Gamma$(MeV)   & $0.04^{+0.01}_{-0.01}$
    & $0.03^{+0.00}_{-0.01}$  \\
\botrule
\end{tabular}
\end{table}

In this manner, the observed ordering of the $1^-$–$2^-$ spin-doublet states clearly reflects the underlying $N\Xi$ interactions. As shown in Table I, the decay widths $\Gamma$ for these states are relatively small, of  the order of 20–40 keV. 
These narrow widths arise from the fact that the $N\Xi$–$\Lambda\Lambda$ coupling interaction in the HAL QCD potential \cite{HALQCD:2019wsz} is localized at short distances, leading to only a weak coupling at low energies.

These level structures in the $\XiNaa$ system for both $T=1$ and $T=0$ isospin channels can be experimentally investigated by producing the system through the $(K^-, K^+)$ and $(K^-, K^0)$ reactions on a $^{10}$B target, respectively. If the ordering of the $1^-$ and $2^-$ levels can be experimentally determined, it will provide valuable constraints on the four spin-isospin components of the $N\Xi$ interaction: $^{31}$S$_0$, $^{33}$S$_1$, $^{11}$S$_0$, and $^{13}$S$_1$.

\section{Summary}\label{sec5}

Thanks to recent advances in theoretical formalism, as well as the development of supercomputers such as "K" and "Fugaku", first-principles determinations of baryon-baryon interactions have become feasible through lattice QCD simulations with
 the HAL QCD method.
In parallel, many-body computational methods have significantly progressed with the aid of high-performance computing. In particular, the Gaussian Expansion Method (GEM) has been successfully applied to systems as complex as the five-body problem.

By combining the state-of-the-art HAL QCD $\XiN$ interactions with GEM, it is now possible to make theoretical predictions for the level structures of $\Xi$-hypernuclei prior to experimental observation. As illustrative examples, we have investigated a variety of systems including $\XiNN$, $\XiNNN$, $\XiNa$, and $\XiNaa$.
Among these, we emphasize that the lightest bound $\Xi$-hypernucleus is predicted to be the four-body $\XiNNN$ system, which can be experimentally produced via the $(K^-, K^0)$ reaction on a $^4$He target at the J-PARC facility.

To extract detailed information on the spin-isospin dependence of the $\XiN$ interaction, we propose to focus on the level structure of the $1^-$ and $2^-$ states in both $T=0$ and $T=1$ channels of the $\XiNaa$ system. In this context, we also propose dedicated experiments to produce this system using $(K^-, K^+)$ and $(K^-, K^0)$ reactions on a $^{10}$B target.
Such measurements at J-PARC are highly anticipated and will provide crucial insights into hyperon-nucleon interactions.

\section*{Acknowledgments}
The present work is supported by 
Grant-In-Aid for Scientific Research on Innovative Areas
'Clustering as a window on the hierarchical structure of quantum systems' (18H05407).
T.D. thanks members of the HAL QCD Collaboration 
for providing lattice QCD results and for fruitful collaborations based on which this paper is prepared.
T.D. was supported in part by Japan Science and Technology Agency (JST) as part of
Adopting Sustainable Partnerships for Innovative Research Ecosystem (ASPIRE), Grant No. JPMJAP2318.

This work was supported in part by 
JSPS Grant-in-Aid for Scientific Research
(Nos. JP18H05236, JP19K03879, JP20H00155, JP23H05439),
and by
"Program for Promoting Researches on the Supercomputer Fugaku" (Simulation for basic science: from fundamental laws of particles to creation of nuclei) 
and (Simulation for basic science: approaching the new quantum era)
(Grants No. JPMXP1020200105, JPMXP1020230411), 
and Joint Institute for Computational Fundamental Science (JICFuS).


\end{document}